\documentclass[12pt]{article}
\usepackage{epsfig,amsmath,amssymb,mathrsfs}
\usepackage[utf8]{inputenc}
\usepackage[colorlinks=true,citecolor=blue,,linktocpage=true,linkcolor=blue,urlcolor=black]{hyperref}
\usepackage{mathtools,slashed}
\usepackage{array}

\usepackage[force]{feynmp-auto}
\usepackage{caption}
\usepackage{subcaption}

\usepackage{cleveref}

  {\end{list}}%

\makeatletter
\renewcommand{\section}{\setcounter{equation}{0}\@startsection
 {section}%
 {1}%
 {0pt}%
 {-1\baselineskip}%
 {0.4\baselineskip}%
 {\bfseries\large}}%
\renewcommand{\subsection}{\@startsection
 {subsection}%
 {2}%
 {0pt}%
 {-0.75\baselineskip}%
 {0.2\baselineskip}%
 {\bfseries}}%
\renewcommand{\subsubsection}{\@startsection
 {subsubsection}%
 {3}%
 {0pt}%
 {-0.5\baselineskip}%
 {0.1\baselineskip}%
 {\sc}}%
\makeatother

\DeclareMathAlphabet{\mathpzc}{OT1}{pzc}{m}{it}

\def\be{\begin{equation}}
\def\ee{\end{equation}}
\def\a{\alpha}

\def\g5{\gamma_{5}}

\def\la{\lambda}

\def\m{\mu}
\def\n{\nu}
\def\r{\rho}
\def\s{\sigma}

\def\idxn{\int\!\! d^{n}\!x}

\newcommand{\bea}{\begin{eqnarray}}
\newcommand{\eea}{\end{eqnarray}}
\newcommand{\beann}{\begin{eqnarray*}}
\newcommand{\eeann}{\end{eqnarray*}}
\newcommand{\ba}{\begin{array}}
\newcommand{\ea}{\end{array}}

 \def\g {\gamma}

\newcommand{\email}[1]{\href{mailto:#1}{\tt #1}}

\begin{document}

\rightline{\scriptsize{FTUAM-17-28}\quad\scriptsize{IFT-UAM/CSIC-17-115} \quad\scriptsize{FTI/UCM 37-2017}}
\vglue 50pt

\begin{center}

{\LARGE \bf Unimodular Gravity and General Relativity UV divergent contributions to the scattering of massive scalar particles}\\
\vskip 1.0true cm
{\Large S. Gonzalez-Martin}$^{\dagger}$ {\large and} {\Large C. P. Martin}$^{\ddag}$
\\
\vskip .7cm
{
	$^{\dagger}$Departamento de F\'isica Te\'orica and Instituto de F\'{\i}sica Te\'orica (IFT-UAM/CSIC),\\
	Universidad Aut\'onoma de Madrid, Cantoblanco, 28049, Madrid, Spain\\
	\vskip .1cm
	$^{\ddag}${Universidad Complutense de Madrid (UCM), Departamento de Física Teórica I, Facultad de Ciencias Físicas,  Av. Complutense S/N (Ciudad Univ.), 28040 Madrid, Spain}
	
	\vskip .5cm
	\begin{minipage}[l]{.9\textwidth}
		\begin{center}
			\textit{E-mail:}
			\email{sergio.gonzalezm@uam.es},
			\email{carmelop@fis.ucm.es}
			
		\end{center}
	\end{minipage}
}
\end{center}
\thispagestyle{empty}

\begin{abstract}
We work out the one-loop and order $\kappa^2 m_\phi^2$ UV divergent contributions, coming from Unimodular Gravity and General Relativity, to the
S matrix element of the scattering process $\phi + \phi\rightarrow \phi + \phi$ in a $\lambda \phi^4$ theory with mass $m_\phi$. We show that both Unimodular Gravity and General Relativity give rise to the same UV divergent contributions in Dimension Regularization. This seems to be  at odds with the known result that in a multiplicative MS dimensional regularization scheme the General Relativity corrections, in the de Donder gauge, to  the beta function $\beta_{\lambda}$ of the $\lambda$ coupling do not vanish, whereas the Unimodular Gravity corrections, in a certain gauge, do vanish. Actually, we show that the UV divergent contributions to the 1PI Feynman diagrams which give rise to those non-vanishing corrections to
$\beta_{\lambda}$ do not contribute to the UV divergent behaviour of the S matrix element of $\phi + \phi\rightarrow \phi + \phi$ and this shows that any physical consequence --such existence of asymptotic freedom due to gravitational interactions-- drawn from the value of
$\beta_{\lambda}$ is not physically meaningful.

\end{abstract}

\vfill
\clearpage

\section{Introduction}

In Unimodular Gravity  the vacuum energy does not gravitate. Actually, when Unimodular Gravity is coupled to matter there is no term in the classical action where the graviton field is coupled to the potential. Thus, a Wilsonian solution of the problem that arises in General Relativity of the huge disparity between the actual value of the Cosmological Constant and its theoretically expected value seems to  show up \cite{Weinberg:1988cp, Henneaux:1989zc, Smolin:2009ti}.

At the classical level  Unimodular Gravity and General Relativity are equivalent theories \cite{Ellis:2010uc, Ellis:2013eqs, Gao:2014nia, Cho:2014taa}, at least as far as the classical equations of motion can tell  \cite{Oda:2016knt, Oda:2016nvc, Chaturvedi:2016fea}. Putting aside the matter of the Cosmological Constant problem mentioned above, whether such equivalence survives the quantization process  is still an open issue; even for physical phenomena where the Cosmological Constant can be effectively set to zero. Several papers have been published where this quantum
equivalence has been discussed: see Refs. \cite{Alvarez:2005iy, Kluson:2014esa, Saltas:2014cta,Eichhorn:2015bna, Alvarez:2015sba, Bufalo:2015wda, Alvarez:2016uog, Martin:2017ewb, Gonzalez-Martin:2017bvw, Ardon:2017atk}. However, only in two of them  \cite{Martin:2017ewb, Gonzalez-Martin:2017bvw} the coupling of Unimodular Gravity with matter has been considered.

In Ref. \cite{Gonzalez-Martin:2017bvw} the coupling of Unimodular Gravity to a massive $\lambda\phi^4$ theory was introduced and the corrections to the beta function of the coupling $\lambda$ coming from Unimodular Gravity were computed. The results obtained point in the direction that, when coupled to the $\lambda\phi^4$ theory,  Unimodular Gravity and General Relativity are equivalent  at the quantum level, at least when the Cosmological Constant can be dropped and for the one-loop UV divergent behaviour considered. However, this conclusion regarding the UV behaviour of these theories --General Relativity plus $\lambda\phi^4$ and Unimodular Gravity coupled to $\lambda\phi^4$-- cannot be considered as final since, as shown in Ref. \cite{Gonzalez-Martin:2017bvw}, the gravitational corrections to the beta function of the coupling $\lambda$ have a very dubious physical meaning. To settle this issue for once and all is important since it has been argued \cite{Rodigast:2009zj, Pietrykowski:2012nc} that the General Relativity corrections to the beta function of the coupling $\lambda$ gives rise to asymptotic freedom, with obvious implications on the Higgs physics.

The purpose of this paper is to compute the one-loop and order $\kappa^2 m_\phi^2$ UV divergent  contributions to the S matrix element of the scattering process $\phi + \phi\rightarrow \phi + \phi$ in a massive --with mass $m_\phi$,  $\lambda \phi^4$-- theory coupled either to General Relativity or to Unimodular Gravity, both in the vanishing Cosmological Constant situation. We shall show that such UV divergent behaviour is  the same in Unimodular Gravity case as in the General Relativity instance and  this is in spite of the fact this equivalence does not hold  Feynman diagram by Feynman diagram. This result is not trivial since  Unimodular Gravity does not couple to the scalar potential in the classical action and it provides further evidence that Unimodular Gravity and General Relativity are equivalent at the quantum level and for zero Cosmological Constant. As a side result, we shall show that the UV divergent contributions which give rise to the non-vanishing gravitational corrections to the beta function of the coupling $\lambda$ computed in \cite{ Gonzalez-Martin:2017bvw,Rodigast:2009zj, Pietrykowski:2012nc} are  completely useless for characterizing UV divergent behaviour S matrix element of the $\phi + \phi\rightarrow \phi + \phi$ scattering.
This is completely at odds with the non-gravitational corrections to the beta function of $\lambda$ and it shows beyond the shadow of a doubt that the gravitational corrections to the beta function of the coupling constants lack, in general, any intrinsic physical meaning. This also applies to the physical implications of a beta function turning negative due to the gravitational corrections.

The lay out of this paper is as follows. In section 2 we display the relevant formulae that are needed to carry out the computations in section 3.
Section 3 is devoted to the computation the one-loop and order $\kappa^2 m_\phi^2$, $m_\phi$ being the mass of the scalar particle,  UV divergent contributions to the S matrix element of the scattering $\phi + \phi\rightarrow \phi + \phi$. Finally, we have a section to discuss the results
presented in the paper.

\section{ Gravity coupled to $\lambda\phi^4$}

In this section we shall just display the classical actions of General Relativity and Unimodular Gravity coupled to $\lambda \phi^4$ and the graviton free propagator in each case.

\subsection{General Relativity coupled to $\lambda\phi^4$}

It goes without saying that the classical action of General relativity coupled to $\lambda \phi^4$ reads
\begin{equation}
\begin{array}{l}
{S_{GR\phi}\;=\;S_{EH}\,+\,S^{(GR)}_{\lambda\phi^4}}\\[8pt]
{\displaystyle S_{EH}=-\dfrac{2}{\kappa^2}\idxn\;\sqrt{-g} R[g_{\mu\nu}]}\\[8pt]
{\displaystyle S^{(GR)}_{\lambda\phi^4}\,=\,\idxn\,\sqrt{-g}\Big[\dfrac{1}{2}g^{\mu\nu}\partial_\mu\phi\partial_\nu\phi-\dfrac{1}{2}M^2\phi^2-\dfrac{\lambda}{4!}\phi^4\Big]}
\end{array}
\label{classactionGR}
\end{equation}
where $\kappa^2=32\pi G$, $R[g_{\mu\nu}]$ is the scalar curvature for the metric $g_{\mu\nu}$.

Using the standard splitting
\begin{equation}
g_{\mu\nu}=\eta_{\mu\nu}\,+\,\kappa h_{\mu\nu};
\label{gsplittingGR}
\end{equation}
and the generalized de Donder gauge-fixing term
\begin{equation*}
\idxn\,\alpha (\partial^\mu h_{\mu\nu}-\partial_\nu h)^2,\quad h=h_{\mu\nu}\eta^{\mu\nu},
 \end{equation*}
which depends on the gauge parameter $\alpha$, one obtains the following free propagator of the graviton field $h_{\mu\nu}$:
\begin{equation}
\begin{array}{l}
{\langle h_{\mu\nu}(k)h_{\rho\sigma}(-k)\rangle=
\dfrac{i}{2k^2}\left(\eta_{\m\s}\eta_{\n\r}+\eta_{\m\r}\eta_{\n\s}-\eta_{\m\n}\eta_{\r\s}\right)}\\[8pt]
{\phantom{\langle h_{\mu\nu}(k)h_{\rho\sigma}(-k)\rangle=}
-\dfrac{i}{(k^2)^2}\,\left(\dfrac{1}{2}+\alpha\right)\left(\eta_{\m\r}k_\n k_\s+\eta_{\m\r}k_\n k_\s+\eta_{\n\r}k_\m k_\s+\eta_{\n\s}k_\m k_\r\right)}.
\end{array}
\label{GRpropagator}
\end{equation}	
$\eta^{\mu\nu}$ denotes the Minkowski, $(+,-,-,-)$, metric.

Up to first order in $\kappa$, $S^{(GR)}_{\lambda\phi^4}$ in (\ref{classactionGR}) is given by
\begin{equation}
S^{(GR)}_{\lambda\phi^4}=\,\idxn\,\Big[\dfrac{1}{2}\partial_\mu\phi\partial^\mu\phi-\dfrac{1}{2}M^2\phi^2-\dfrac{\lambda}{4!}\phi^4-\dfrac{\kappa}{2}\,T^{\mu\nu}h_{\mu\nu}\Big]+O(\kappa^2),
\label{Sphi4GR}
\end{equation}
where
 \begin{equation*}
 T^{\mu\nu}= \partial^{\mu}\phi\partial^{\nu}\phi-\eta^{\mu\nu}\big(\dfrac{1}{2}\partial_\la\phi\partial^\la\phi-\dfrac{1}{2}M^2\phi^2-\dfrac{\lambda}{4!}\phi^4\big).
 \end{equation*}
In \eqref{Sphi4GR}, contractions are carried out with $\eta_{\mu\nu}$, the Minkowski metric.

\subsection{Unimodular Gravity coupled to $\lambda\phi^4$}

Let $\hat{g}_{\mu\nu}$ denote the Unimodular --ie, with determinant equal to $(-1)$-- metric of the $n$ dimensional spacetime manifold. We shall assume the mostly minus signature for the metric. Then,
the classical action of Unimodular gravity coupled to  $\lambda\phi^4$ reads
\begin{equation}
\begin{array}{l}
{S_{UG\phi}\;=\;S_{UG}\,+\,S^{(UG)}_{\lambda\phi^4}}\\[8pt]
{S_{UG}=-\dfrac{2}{\kappa^2}\idxn\; R[\hat{g}_{\mu\nu}]}\\[8pt]
{S^{(UG)}_{\lambda\phi^4}\,=\,\idxn\,\Big[\dfrac{1}{2}\hat{g}^{\mu\nu}\partial_\mu\phi\partial_\nu\phi-\dfrac{1}{2}M^2\phi^2-\dfrac{\lambda}{4!}\phi^4\Big]}
\end{array}
\label{classactionUG}
\end{equation}
where $\kappa^2=32\pi G$, $R[\hat{g}_{\mu\nu}]$ is the scalar curvature for the unimodular metric.

To quantize the theory we shall proceed as in Refs.~\cite{ Alvarez:2005iy, Alvarez:2015sba,Alvarez:2006uu} and  introduce the unconstrained fictitious metric, $g_{\mu\n}$, thus
\begin{equation}
\hat{g}_{\mu\nu}=(-g)^{-1/n}\,g_{\mu\nu};
\label{fictitious}
\end{equation}
where $g$ is the determinant of $g_{\mu\nu}$. Then, we shall express the action in (\ref{classactionGR}) in terms of the fictitious metric $g_{\mu\nu}$ by using  (\ref{fictitious}). Next,  we shall
split  $g_{\mu\nu}$ as in \eqref{gsplittingGR}
\begin{equation}
g_{\mu\nu}=\eta_{\mu\nu}\,+\,\kappa h_{\mu\nu};
\label{gsplitting}
\end{equation}
and, finally, we shall defined the path integral  by integration over $h_{\mu\nu}$ and the matter fields, once an appropriate BRS invariant action  has been constructed.

Since our computations will always involve the scalar field $\phi$ and will be of order $\kappa^2$, we only need --as will become clear in the sequel-- the free propagator of $h_{\mu\nu}$,
$\langle h_{\mu\nu}(k)h_{\rho\sigma}(-k)\rangle$,  and the expansion of $S_{\lambda\phi^4}$ up to first order in $\kappa$. Using the gauge-fixing procedure discussed in Ref.~\cite{Alvarez:2015sba}, one obtains
\begin{equation}
\begin{array}{l}
{\langle h_{\mu\nu}(k)h_{\rho\sigma}(-k)\rangle=
\dfrac{i}{2k^2}\left(\eta_{\m\s}\eta_{\n\r}+\eta_{\m\r}\eta_{\n\s}\right)-\dfrac{i}{k^2}\dfrac{\a^2n^2-n+2}{\a^2 n^2(n-2)}\eta_{\m\n}\eta_{\r\s}}\\[8pt]
{\phantom{\langle h_{\mu\nu}(k)h_{\rho\sigma}(-k)\rangle=}
+\dfrac{2i}{n-2}\left(\dfrac{k_\r k_\s \eta_{\m\n}}{(k^2)^2}+\dfrac{k_\m k_\n \eta_{\r\s}}{(k^2)^2}\right)-\dfrac{2in}{n-2}\dfrac{k_{\m}k_{\n}k_{\r}k_{\s}}{(k^2)^3}.
}
\end{array}
\label{propagatorug}
\end{equation}	
The expansion of $S_{\lambda\phi^4}$ in powers of $\kappa$ reads
\begin{equation}
S^{(UG)}_{\lambda\phi^4}=\,\idxn\,\Big[\dfrac{1}{2}\partial_\mu\phi\partial^\mu\phi-\dfrac{1}{2}M^2\phi^2-\dfrac{\lambda}{4!}\phi^4-\dfrac{\kappa}{2}\,T^{\mu\nu}\hat{h}_{\mu\nu}\Big]+O(\kappa^2),
\label{Sphi4expan}
\end{equation}
where $\hat{h}_{\mu\nu}=h_{\mu\nu}-\dfrac{1}{n}h$, with $h=\eta_{\mu\nu}h^{\mu\nu}$, is the traceless part of $h_{\mu\nu}$ and
 \begin{equation}
 T^{\mu\nu}= \partial^{\mu}\phi\partial^{\nu}\phi-\eta^{\mu\nu}\big(\dfrac{1}{2}\partial_\mu\phi\partial^\mu\phi-\dfrac{\lambda}{2}M^2\phi^2-\dfrac{\lambda}{4!}\phi^4\big).
 \label{EMphi}
 \end{equation}
Again, the contractions in \eqref{Sphi4expan} are carried out with the Minkowski metric $\eta_{\mu\nu}$.

Notice that the summand in $T^{\mu\nu}$ which is proportional to $\eta^{\mu\nu}$ does not actually contribute to  $T^{\mu\nu}\hat{h}_{\mu\nu}$, since $\hat{h}_{\mu\nu}$ is traceless. In terms of Feynman diagrams, this amounts to saying that the $\eta^{\mu\nu}$ part of $T^{\mu\nu}$ will never
contribute to a given diagram since it will always be contracted with a free propagator involving $\hat{h}_{\mu\nu}$. This as opposed to the case of General Relativity and makes the agreement between  General Relativity coupled to matter and Unimodular Gravity coupled to matter quite surprising already at the one-loop level.

It is the free propagator of $\hat{h}_{\mu\nu}$, $\langle \hat{h}_{\mu\nu}(k)\hat{h}_{\rho\sigma}(-k)\rangle$, and not the full graviton propagator in (\ref{propagatorug}), the correlation function that will enter the computations carried out in this paper. From (\ref{propagatorug}) one readily obtains that
\begin{equation}
\begin{array}{l}
{\langle\hat{h}_{\mu\nu}(k)\hat{h}_{\rho\sigma}(-k)\rangle=\dfrac{i}{2k^2}\left(\eta_{\m\s}\eta_{\n\r}+\eta_{\m\r}\eta_{\n\s}-\dfrac{2}{n-2}\eta_{\m\n}\eta_{\r\s}\right)+}\\[8pt]
{\phantom{\langle\hat{h}_{\mu\nu}(k)\hat{h}_{\rho\sigma}(-k)\rangle=}
\dfrac{2i}{n-2}\dfrac{k_{\mu}k_{\nu}\eta_{\rho\sigma}+k_{\rho}k_{\sigma}\eta_{\mu\nu}}{(k^2)^2}-
\dfrac{2in}{n-2}\dfrac{k_{\mu}k_{\nu}k_{\rho}k_{\sigma}}{(k^2)^3}.}
\label{tracelessprop}
\end{array}
\end{equation}

\section{The $\phi + \phi\rightarrow \phi + \phi $ scattering at one-loop and at order $\kappa^2 m_\phi^2$ }

The purpose of this section is to work out the one-loop and order $\kappa^2 m_\phi^2$ UV divergent contribution, coming from General Relativity and Unimodular Gravity, to the dimensionally regularized S-matrix element of the  $\phi + \phi\rightarrow \phi + \phi $ scattering process and discuss the
meaning of the results we shall obtain.

\subsection{General Relativity contributions}

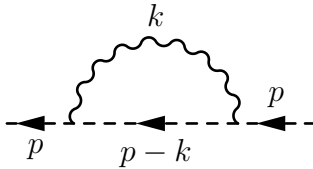
\begin{figure}[h]
	\centering
	
	\begin{fmffile}{scalarprop}
		\begin{fmfgraph*}(120,60)
			\fmfleft{i}
			\fmfright{o}
			\fmf{scalar,tension=3,label=$p$}{v1,i}
			\fmf{scalar,tension=0.7,label=$p-k$}{v2,v1}
			\fmf{scalar,tension=3,label=$p$}{o,v2}
			\fmf{wiggly,left,tension=0.7,label=$k$}{v1,v2}
		\end{fmfgraph*}\vspace{-0.4cm}
	\end{fmffile}
	\caption{$i\Gamma_{\phi\phi}(p^2;\kappa)$}
	\label{scalarprop}
\end{figure}

Let us consider the General Relativity case in the first place.  To define the S-matrix of the $\phi + \phi\rightarrow \phi + \phi  $ scattering at one-loop, we need the one-loop propagator of the scalar field $\phi$ to have simple pole at the physical mass, $m_{\phi}$ with residue $i$. This is achieved by introducing the following mass and wave function renormalizations
\begin{equation}
\begin{array}{l}
{m_\phi^2=M^2+i\Gamma_{\phi\phi}(p^2=m_\phi^2, \kappa)}\\[8pt]
{\phi=\phi_R\big[1-i\Gamma^{'}_{\phi\phi}(p^2=m_\phi^2, \kappa)\big]^{-1/2},\quad \Gamma^{'}_{\phi\phi}(p^2,\kappa)=\dfrac{\partial \Gamma_{\phi\phi}(p^2, \kappa)}{\partial p^2} },
\end{array}
\label{residuer}
\end{equation}
where $M^2$ and $\phi$ are the bare objects in the action in \eqref{classactionGR}. In the previous equation, the symbol $\Gamma_{\phi\phi}(p^2)$ denotes the one-loop contribution to the 1PI two-point function of the scalar field. The General Relativity contribution, $i\Gamma^{(GR)}_{\phi\phi}(p^2, \kappa)$ --the non-gravitational ones can be found in standard textbooks-- to  $i\Gamma_{\phi\phi}(p^2)$ is given by the diagram in Figure \ref{scalarprop} and it reads
\begin{equation}
i\Gamma^{(GR)}_{\phi\phi}(p^2,\kappa)= \Big(\dfrac{1}{16\pi^2\epsilon}\Big)\left[1+ \left(\dfrac{1}{2}+\alpha\right)\right]\kappa^2 M^2 (p^2-M^2)\,+\,\text{UV finite contributions},
\label{GR1PI2pt}
\end{equation}
where $n=4+2\epsilon$ is the spacetime dimension. The wavy line in Figure \ref{scalarprop} denotes the free propagator in (\ref{GRpropagator}).

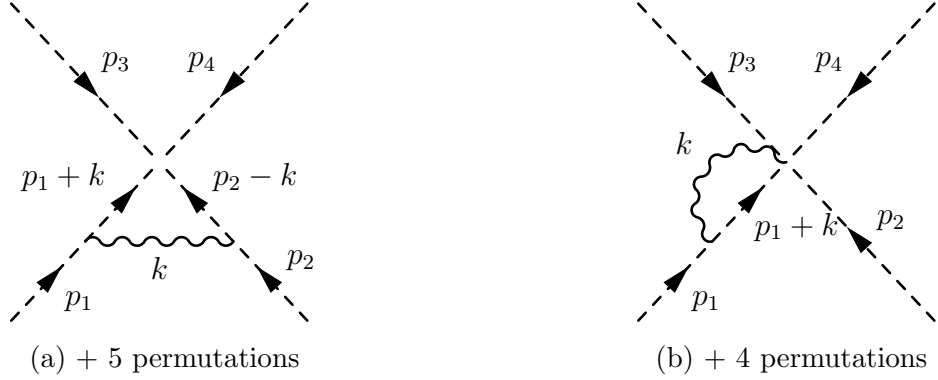
\begin{figure}[!h]
	\centering
	\begin{subfigure}[t]{0.3\linewidth}
		\begin{fmffile}{phi41}
			\begin{fmfgraph*}(140,120)
				\fmfleft{i1,i2}
				\fmfright{o1,o2}
				\fmf{scalar,tension=0.5,label=$p_3$}{i2,v2}
				\fmf{scalar,tension=1,label=$p_1$}{i1,v1}
				\fmf{scalar,tension=1,label=$p_1+k$,l.side=left}{v1,v2}
				\fmf{scalar,tension=0.5,label=$p_4$}{o2,v2}
				\fmf{scalar,tension=1,label=$p_2$,l.side=right}{o1,v3}
				\fmf{scalar,tension=1,label=$p_2-k$,l.side=right}{v3,v2}
				\fmf{wiggly,tension=0,label=$k$,l.side=left}{v3,v1}	
			\end{fmfgraph*}
		\end{fmffile}
		\caption{+ 5 permutations}
	\end{subfigure}
	\hspace{3cm}
	\begin{subfigure}[t]{0.3\linewidth}
		\begin{fmffile}{phi42}
			\begin{fmfgraph*}(140,120)
				\fmfleft{i1,i2}
				\fmfright{o1,o2}
				\fmf{scalar,tension=0.5,label=$p_3$}{i2,v2}
				\fmf{scalar,tension=1,label=$p_1$}{i1,v1}
				\fmf{scalar,tension=1,label=$\hspace{-1.5mm}p_1+k$,l.side=right}{v1,v2}
				\fmf{scalar,tension=0.5,label=$p_4$}{o2,v2}
				\fmf{scalar,tension=0.5,label=$p_2$,l.side=right}{o1,v2}
				\fmf{wiggly,left,tension=0,label=$k$,l.side=left}{v1,v2}	
			\end{fmfgraph*}
		\end{fmffile}
		\caption{+ 4 permutations}
		\label{2b}
	\end{subfigure}
	\caption{1 loop scalar four-point function: $i\Gamma^{(GR)\,\rm{or}\, (UG)}_{\phi\phi\phi\phi}(p_1,p_2,p_3,p_4;\kappa)$}
	\label{scalar4p}
\end{figure}

\begin{figure}[h!]
	\centering
	
	\centering
	\begin{fmffile}{phi4count}
		\begin{fmfgraph*}(140,140)
			\fmfbottom{i,o}
			\fmftop{t,t1}
			\fmf{scalar,tension=1,label=$p_1$}{t,v1}
			\fmf{scalar,tension=1,label=$p_2$}{t1,v1}
			\fmf{scalar,tension=1,label=$p_3$}{i,v1}
			\fmf{scalar,tension=1,label=$p_4$}{o,v1}
			\fmfv{decor.shape=circle,decor.filled=empty,
				decor.size=(.2w)}{v1}
			\fmfv{l=\scalebox{3.5}{$\mathbf {\times}$},label.dist=0}{v1}
		\end{fmfgraph*}
	\end{fmffile}
	\caption{Onshell counterterm}
	\label{phi4count}
\end{figure}
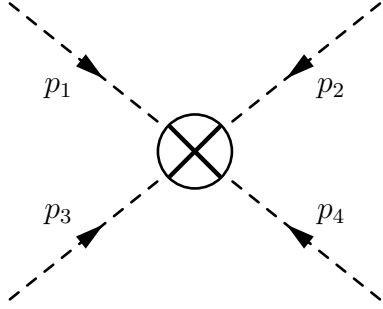

\begin{figure}[ht!]
    \centering
    \begin{subfigure}{\linewidth}
        \centering
    \begin{fmffile}{non1pi1}
        \begin{fmfgraph*}(160,120)
            \fmfleft{i,i1}
            \fmfright{f,f1}
            \fmf{scalar,tension=0.5,label=$p_1$,l.side=left}{i,v1}
            \fmf{scalar,tension=0.5,label=$p_2$}{i1,v1}
            \fmf{wiggly,tension=0.7}{v1,v2}
            \fmf{scalar,right,tension=0.25,label=$k$}{v3,v2}
\fmf{scalar,right,tension=0.25,label=$k+p_1+p_2$}{v2,v3}
            \fmf{scalar,tension=0.5,label=$p_3$,l.side=left}{v3,f}
\fmf{scalar,tension=0.5,label=$p_4$,l.side=right}{v3,f1}
        \end{fmfgraph*}
    \end{fmffile}
    \begin{fmffile}{non1pi2}
        \begin{fmfgraph*}(160,120)
            \fmfleft{i,i1}
            \fmfright{f,f1}
            \fmf{scalar,tension=0.5,label=$p_1$,l.side=left}{i,v1}
            \fmf{scalar,tension=0.5,label=$p_2$}{i1,v1}
            \fmf{wiggly,tension=0.7}{v2,v3}
            \fmf{scalar,right,tension=0.25,label=$k$}{v2,v1}
\fmf{scalar,right,tension=0.25,label=$k+p_1+p_2$}{v1,v2}
            \fmf{scalar,tension=0.5,label=$p_3$,l.side=left}{v3,f}
            \fmf{scalar,tension=0.5,label=$p_4$}{v3,f1}
        \end{fmfgraph*}
    \end{fmffile}
\label{scalarn1p12}
\end{subfigure}

\begin{subfigure}{\linewidth}
    \vspace{0.5cm}
    \caption*{+ u and t channels}
    \end{subfigure}

    \caption{Non-1PI diagrams}
    \label{scalarn1p1}
\end{figure}
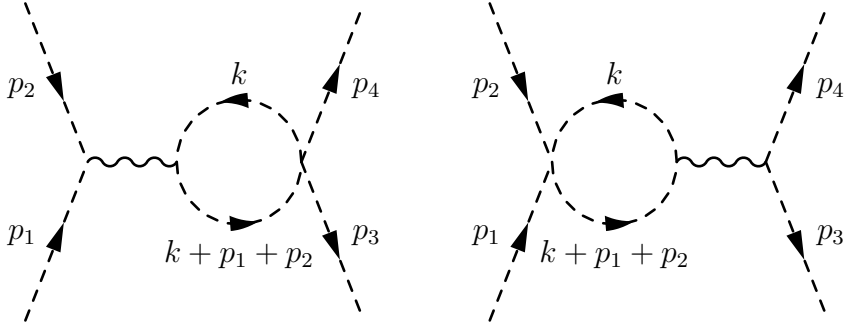

Now, in terms of the $m_\phi$ and $\phi_r$ defined in (\ref{residuer}), the one-loop and order $\kappa^2 m_\phi^2$ General Relativity contribution to the dimensionally regularized S-matrix element of the scattering process $\phi + \phi\rightarrow \phi + \phi$ is given by the sum the diagrams in Figures \ref{scalar4p}, \ref{phi4count} and \ref{scalarn1p1} --bear in mind that the wavy lines represent free propagator in (\ref{GRpropagator}). Notice that the diagram in Figure 3\ref{phi4count} comes from the wave function renormalization in (\ref{residuer}), which guarantees that asymptotically $\phi_r$ is the free field at $t=\pm \infty$. It can be shown that the sum of all the diagrams in
Fig. \ref{scalar4p} is given by
\begin{equation}
\begin{array}{l}
{\displaystyle i\Gamma^{(GR)}_{\phi\phi\phi\phi}(p_1,p_2,p_3,p_4;\kappa)\rvert_{p_i^2=m_\phi^2}=}\\[8pt]
={\displaystyle\Big(\dfrac{-1}{16\pi^2\epsilon}\Big)\left[1+ \left(\dfrac{1}{2}+\alpha\right)\right]\kappa^2 \lambda \Big(\sum_{i<j}\,p_i\cdot p_j\rvert_{p_i^2=m_\phi^2}+4 m_ \phi ^2\Big)\,+\,\text{UV finite contributions}=}\\[8pt]
={\Big(\dfrac{-1}{16\pi^2\epsilon}\Big)\left[1+ \left(\dfrac{1}{2}+\alpha\right)\right]\kappa^2 m_\phi^2 \lambda \big[2\big]\,+\,\text{UV finite contributions}.}
\end{array}
\label{GR4pt1PI}
\end{equation}
Note that $i,j=1,2,3$ and $4$.

Taking into account (\ref{residuer}) and (\ref{GR1PI2pt}), one concludes that contribution to the dimensionally regularized S-matrix coming from the diagram in Figure \ref{phi4count} reads
\begin{equation}
\begin{array}{l}
{i\Gamma^{(GR,ct)}_{\phi\phi\phi\phi}(p_1,p_2,p_3,p_4;\kappa)= \lambda\,\left[1-i\dfrac{\partial\Gamma^{(GR)}_{\phi\phi}}{\partial p^2}(p^2=m_\phi^2, \kappa)\right]^{-2}-\lambda }=\\[8pt]
={\Big(\dfrac{1}{16\pi^2\epsilon}\Big)\left[1+ \left(\dfrac{1}{2}+\alpha\right)\right]\kappa^2 m_\phi^2 \lambda \big[2\big]\,+\,\text{UV finite contributions}.}
\end{array}
\label{GRonshellct}
\end{equation}

From (\ref{GR4pt1PI}) and (\ref{GRonshellct}), one immediately realizes that the
\begin{equation}
i\Gamma^{(GR)}_{\phi\phi\phi\phi}(p_1,p_2,p_3,p_4;\kappa)\rvert_{p_i^2=m_\phi^2}+i\Gamma^{(GR,ct)}_{\phi\phi\phi\phi}(p_1,p_2,p_3,p_4;\kappa) = \text{UV finite contributions},
\label{zero1PI}
\end{equation}
so that the   General Relativity one-loop  and order $\kappa^2 m_\phi^2$ UV divergent contributions to the S-matrix of the process $\phi + \phi\rightarrow \phi + \phi$ may only come from the non-1PI diagrams in Figure \ref{scalarn1p1}. This sum reads
\begin{equation}
\begin{array}{l}
{\displaystyle i{\rm N}\Gamma_{\phi\phi\phi\phi}^{(GR)}(p_1,p_2,p_3,p_4;\kappa)\rvert_{p_i^2=m_\phi^2}=}\\[8pt]
={\Big(\dfrac{-1}{16\pi^2\epsilon}\Big)\left(\dfrac{1}{12}\right)\kappa^2\lambda\left[\dfrac{1}{2}(s+t+u)\rvert_{p_i^2=m_\phi^2}+m_\phi^2\right]\,+\,\text{UV finite contributions}=}\\[8pt]
={\Big(\dfrac{-1}{16\pi^2\epsilon}\Big)\left(\dfrac{1}{2}\right)\kappa^2 m_\phi^2\lambda\,+\,\text{UV finite contributions}.}
\end{array}
\label{N1PIGR}
\end{equation}

We are now ready to display the one-loop and order $\kappa^2 m_\phi^2$ UV contribution to the dimensional regularized S matrix element of the scattering process $\phi+\phi\rightarrow \phi+\phi$ coming from General Relativity. The contribution in question
is obtained by adding the UV divergent contributions in (\ref{GR4pt1PI}), (\ref{GRonshellct}) and (\ref{N1PIGR}) and it reads
\begin{equation}
\Big(\dfrac{-1}{16\pi^2\epsilon}\Big)\left(\dfrac{1}{2}\right)\kappa^2 m_\phi^2\lambda.
\label{UVSMATGR}
\end{equation}
Let us insist on the fact that the contribution in (\ref{UVSMATGR}) only comes from the diagrams in Figure \ref{scalarn1p1}, which are one particle reducible, for the contribution coming from the 1PI diagram in Figure \ref{phi4count} cancels the contributions coming from the diagrams in Figure \ref{scalar4p}, as seen in (\ref{zero1PI}).

\subsection{Unimodular Gravity contributions}

To compute the one-loop and order $\kappa^2 m_\phi^2$ Unimodular Gravity contributions to the S matrix element giving the $\phi +\phi\rightarrow \phi+\phi$ scattering, one proceeds as in the previous subsection,taking into account that now the wavy lines in the Feynman diagrams in Figures \ref{scalarprop}, \ref{scalar4p}, and \ref{scalarn1p1} stand for the traceless free correlation function in (\ref{tracelessprop}) and that the diagrams in Figure \ref{2b} are zero since they come from the contraction of the $\eta^{\mu\nu}$ bit of $T_{\mu\n}$ in (\ref{EMphi})  and the traceless $\hat{h}_{\mu\nu}$ field. Our computations yield the following
results
\begin{equation}
\begin{array}{l}
{i\Gamma^{(UG)}_{\phi\phi}(p^2,\kappa)= 0\,+\,\text{UV finite contributions}}\\[8pt]
{i\Gamma^{(UG)}_{\phi\phi\phi\phi}(p_1,p_2,p_3,p_4;\kappa)\rvert_{p_i^2=m_\phi^2}=0\,+\,\text{UV finite contributions}}\\[8pt]
{i{\rm N}\Gamma_{\phi\phi\phi\phi}^{(UG)}(p_1,p_2,p_3,p_4;\kappa)\rvert_{p_i^2=m_\phi^2}=}\\[8pt]
={\Big(\dfrac{-1}{16\pi^2\epsilon}\Big)\left(\dfrac{1}{12}\right)\kappa^2\lambda\left[\dfrac{1}{2}(s+t+u)\rvert_{p_i^2=m_\phi^2}+m_\phi^2\right]\,+\,\text{UV finite contributions}=}\\[8pt]
={\Big(\dfrac{-1}{16\pi^2\epsilon}\Big)\left(\dfrac{1}{2}\right)\kappa^2 m_\phi^2\lambda\,+\,\text{UV finite contributions},}
\end{array}
\label{1PIUGpole}
\end{equation}
where $i\Gamma^{(UG)}_{\phi\phi}(p^2,\kappa)$ is give by the diagram in Figure \ref{scalarprop}, $i\Gamma^{(UG)}_{\phi\phi\phi\phi}(p_1,p_2,p_3,p_4;\kappa)$ is the sum of all the diagrams --which are not trivially zero-- in Figure \ref{scalar4p} and $i{\rm N}\Gamma_{\phi\phi\phi\phi}^{(UG)}(p_1,p_2,p_3,p_4;\kappa)$ is the
sum of all the diagrams in Figure \ref{scalarn1p1} and $m_\phi$ is the physical mass of the scalar field $\phi$.

By applying the on-shell definitions in (\ref{residuer})  --ie, now $M^2$ and $\phi$ are the bare objects in the action in (\ref{classactionUG})-- to our case, one concludes that for Unimodular Gravity the diagram in Fig. \ref{phi4count} is given by
\begin{equation}
\begin{array}{l}
{i\Gamma^{(UG,ct)}_{\phi\phi\phi\phi}(p_1,p_2,p_3,p_4;\kappa)=\lambda\,\big[1-i\dfrac{\partial\Gamma^{(GR)}_{\phi\phi}}{\partial p^2}(p^2=m_\phi^2, M^2)\big]^{-2}-\lambda= }\\[8pt]
{\phantom{i\Gamma^{(UG,ct)}_{\phi\phi\phi\phi}(p_1,p_2,p_3,p_4;\kappa)=}
=0\,+\,\text{UV finite contributions}.}
\end{array}
\label{UGonshellct}
\end{equation}

Taking into account the results in (\ref{1PIUGpole}) and (\ref{UGonshellct}) and adding the UV divergent contributions to $i\Gamma^{(UG)}_{\phi\phi\phi\phi}(p_1,p_2,p_3,p_4;\kappa)\rvert_{p_i^2=m_\phi^2}$, $i\Gamma^{(UG,ct)}_{\phi\phi\phi\phi}(p_1,p_2,p_3,p_4;\kappa)$ and
$i{\rm N}\Gamma_{\phi\phi\phi\phi}^{(UG)}(p_1,p_2,p_3,p_4;\kappa)\rvert_{p_i^2=m_\phi^2}$, one obtains the  one-loop and order $\kappa^2 m_\phi^2$ UV contribution to the dimensional regularized S matrix element of the scattering process $\phi+\phi\rightarrow \phi+\phi$ coming from Unimodular Gravity, which runs thus
\begin{equation}
\Big(\dfrac{-1}{16\pi^2\epsilon}\Big)\left(\dfrac{1}{2}\right)\kappa^2 m_\phi^2\lambda.
\label{UVSMATUG}
\end{equation}
This is the same contribution that we obtained in the General Relativity case --see (\ref{UVSMATGR}). Notice, however,  that both  $i\Gamma^{(UG)}_{\phi\phi\phi\phi}(p_1,p_2,p_3,p_4;\kappa)\rvert_{p_i^2=m_\phi^2}$ and $i\Gamma^{(UG,ct)}_{\phi\phi\phi\phi}(p_1,p_2,p_3,p_4;\kappa)$ are UV finite, which is at odds with their General Relativity counterparts in (\ref{GR4pt1PI}), (\ref{GRonshellct}).

\section{Discussion}

To get a more physical understanding of the results presented on the two previous subsections, we shall carry out the standard replacement
\begin{equation*}
\dfrac{1}{\epsilon}\rightarrow -\ln \dfrac{\Lambda^2}{\mu^2}
\end{equation*}
in each UV divergent expression of the previous subsection. This way the UV divergent contributions we have computed  as poles at $\epsilon=0$ are interpreted as the logarithmically  divergent contributions arising from virtual particles moving around the loop with momentum $\Lambda$, with $\Lambda$ being the  momentum cutoff. Thus, (\ref{UVSMATGR}) and (\ref{UVSMATUG}) tell us that the sum of all those contributions are the same in General Relativity as in Unimodular Gravity when the S matrix element of the scattering process $\phi + \phi \rightarrow \phi+\phi$ is computed at one-loop and at order $\kappa^2 m_\phi^2$. This contribution being
 \begin{equation*}
\Big(\dfrac{1}{16\pi^2}\Big)\left(\dfrac{1}{2}\right)\kappa^2 m_\phi^2\lambda\,\ln \dfrac{\Lambda^2}{\mu^2}.
\end{equation*}

 This is a non-trivial result since the way the graviton field $h_{\mu\nu}$, couples to the energy-momentum tensor $T^{\mu\nu}$, in General Relativity is not the same as in Unimodular Gravity --see (\ref{Sphi4GR}) and (\ref{Sphi4expan}). Indeed, in Unimodular Gravity only the traceless part of  $h^{\mu\nu}$ is seen by $T^{\mu\nu}$, which imply that some diagrams --the  in Figure \ref{2b}-- are absent --ie, they vanish off-shell-- for Unimodular Gravity, while they are non zero for General Relativity. What is more, even  the on-shell sum of all the 1PI diagrams in Figure \ref{scalar4p} yield a nonvanishing UV divergent contribution in General Relativity -see (\ref{GR4pt1PI}), whereas the corresponding contribution is zero for Unimodular Gravity.

 Another issue that deserves being  discussed is the following. As shown in Ref. \cite{Rodigast:2009zj} the General Relativity correction to the beta function of the coupling $\lambda$ computed in the  multiplicative MS scheme, applied off-shell,  are not zero. This correction comes
from the UV divergent part of the off-shell sum of all the diagrams in Figure \ref{scalar4p}, upon introducing a multiplicative, and off-shell, MS wave function renormalization of $\phi$. And yet, as we have shown above, the UV divergent bits of the diagrams in Figure \ref{scalar4p} do not contribute to the UV divergent behaviour of S matrix element of the $\phi +\phi\rightarrow \phi + \phi$ scattering. This clearly shows that the beta function in question lacks any intrinsic physical meaning, since it is irrelevant in understanding the UV divergent behaviour of the S matrix element in question. This conclusion is in complete
agreement with the analysis carried out in \cite{Gonzalez-Martin:2017bvw} --see also Ref. \cite{Anber:2010uj}-- where it has been shown that those nonvanishing corrections found in Ref. \cite{Rodigast:2009zj} can be set to zero by considering a non-multiplicative wave function renormalization.

As a final remark, notice that the UV divergent behaviour of the S matrix element that we have computed comes entirely from the one-loop contribution to the 1PI function $\langle h_{\mu\nu}\phi\phi \rangle^{(1PI)}$, which at tree level does not involve $\lambda$. This shows that any physical consequence --such as existence of asymptotic freedom due to gravitational interactions-- drawn from the value of $\beta_{\lambda}$ in not physically meaningful.

\section{Acknowledgements}
We are grateful to E. Alvarez for illuminating discussions.
This work has received funding from the European Unions Horizon 2020 research and innovation programme under the Marie Sklodowska-Curie grants agreement No 674896 and No 690575. We also have been partially supported by by the Spanish MINECO through grants
FPA2014-54154-P and FPA2016-78645-P, COST actions MP1405 (Quantum Structure of Spacetime) and  COST MP1210 (The string theory Universe).  S.G-M acknowledge the support of the Spanish Research Agency (Agencia Estatal de Investigaci\'on) through the grant IFT Centro de Excelencia Severo Ochoa SEV-2016-0597.



\begin{thebibliography}{99}



\bibitem{Weinberg:1988cp}
  S.~Weinberg,
  Rev.\ Mod.\ Phys.\  {\bf 61} (1989) 1.
  doi:10.1103/RevModPhys.61.1

\bibitem{Henneaux:1989zc}
  M.~Henneaux and C.~Teitelboim,
  Phys.\ Lett.\ B {\bf 222} (1989) 195.
  doi:10.1016/0370-2693(89)91251-3


\bibitem{Smolin:2009ti}
  L.~Smolin,
  Phys.\ Rev.\ D {\bf 80} (2009) 084003
  doi:10.1103/PhysRevD.80.084003
  [arXiv:0904.4841 [hep-th]].


\bibitem{Ellis:2010uc}
  G.~F.~R.~Ellis, H.~van Elst, J.~Murugan and J.~P.~Uzan,
  Class.\ Quant.\ Grav.\  {\bf 28} (2011) 225007
  doi:10.1088/0264-9381/28/22/225007
  [arXiv:1008.1196 [gr-qc]].

 \bibitem{Ellis:2013eqs}
  G.~F.~R.~Ellis,
  Gen.\ Rel.\ Grav.\  {\bf 46} (2014) 1619
  doi:10.1007/s10714-013-1619-5
  [arXiv:1306.3021 [gr-qc]].


\bibitem{Gao:2014nia}
  C.~Gao, R.~H.~Brandenberger, Y.~Cai and P.~Chen,
  JCAP {\bf 1409} (2014) 021
  doi:10.1088/1475-7516/2014/09/021
  [arXiv:1405.1644 [gr-qc]].

\bibitem{Cho:2014taa}
  I.~Cho and N.~K.~Singh,
  Class.\ Quant.\ Grav.\  {\bf 32} (2015) no.13,  135020
  doi:10.1088/0264-9381/32/13/135020
  [arXiv:1412.6205 [gr-qc]].






\bibitem{Oda:2016knt}
  I.~Oda,
  Mod.\ Phys.\ Lett.\ A {\bf 32} (2017) no.03,  1750022
  doi:10.1142/S0217732317500225
  [arXiv:1607.06562 [gr-qc]].

\bibitem{Oda:2016nvc}
  I.~Oda,
  Mod.\ Phys.\ Lett.\ A {\bf 31} (2016) no.36,  1650206
  doi:10.1142/S0217732316502060
  [arXiv:1608.00285 [gr-qc]].


\bibitem{Chaturvedi:2016fea}
  P.~Chaturvedi, N.~K.~Singh and D.~V.~Singh,
  International Journal of Modern Physics D,1750082 (2017)
  doi:10.1142/S0218271817500821
  [arXiv:1610.07661 [gr-qc]].


\bibitem{Alvarez:2005iy}
  E.~Alvarez,
  JHEP {\bf 0503} (2005) 002
  doi:10.1088/1126-6708/2005/03/002
  [hep-th/0501146].



\bibitem{Kluson:2014esa}
  J.~Kluson,
  Phys.\ Rev.\ D {\bf 91} (2015) no.6,  064058
  doi:10.1103/PhysRevD.91.064058
  [arXiv:1409.8014 [hep-th]].


\bibitem{Saltas:2014cta}
  I.~D.~Saltas,
  Phys.\ Rev.\ D {\bf 90} (2014) no.12,  124052
  doi:10.1103/PhysRevD.90.124052
  [arXiv:1410.6163 [hep-th]].



\bibitem{Eichhorn:2015bna}
  A.~Eichhorn,
  JHEP {\bf 1504} (2015) 096
  doi:10.1007/JHEP04(2015)096
  [arXiv:1501.05848 [gr-qc]].

\bibitem{Alvarez:2015sba}
  E.~Álvarez, S.~González-Martín, M.~Herrero-Valea and C.~P.~Martín,
  JHEP {\bf 1508} (2015) 078
  doi:10.1007/JHEP08(2015)078
  [arXiv:1505.01995 [hep-th]].





\bibitem{Bufalo:2015wda}
  R.~Bufalo, M.~Oksanen and A.~Tureanu,
  Eur.\ Phys.\ J.\ C {\bf 75} (2015) no.10,  477
  doi:10.1140/epjc/s10052-015-3683-3
  [arXiv:1505.04978 [hep-th]].


\bibitem{Alvarez:2016uog}
  E.~Alvarez, S.~Gonzalez-Martin and C.~P.~Martin,
  Eur.\ Phys.\ J.\ C {\bf 76} (2016) no.10,  554
  doi:10.1140/epjc/s10052-016-4384-2
  [arXiv:1605.02667 [hep-th]].

\bibitem{Martin:2017ewb}
  C.~P.~Martin,
  JCAP {\bf 1707} (2017) no.07,  019
  doi:10.1088/1475-7516/2017/07/019
  [arXiv:1704.01818 [hep-th]].


\bibitem{Gonzalez-Martin:2017bvw}
  S.~Gonzalez-Martin and C.~P.~Martin,
  Phys.\ Lett.\ B {\bf 773} (2017) 585
  doi:10.1016/j.physletb.2017.09.011
  [arXiv:1707.06667 [hep-th]].

\bibitem{Ardon:2017atk}
  R.~d.~L.~Ardon, N.~Ohta and R.~Percacci,
  arXiv:1710.02457 [gr-qc].



\bibitem{Alvarez:2006uu}
  E.~Alvarez, D.~Blas, J.~Garriga and E.~Verdaguer,
  Nucl.\ Phys.\ B {\bf 756} (2006) 148
  doi:10.1016/j.nuclphysb.2006.08.003
  [hep-th/0606019].



\bibitem{Rodigast:2009zj}
  A.~Rodigast and T.~Schuster,
  Phys.\ Rev.\ Lett.\  {\bf 104} (2010) 081301
  doi:10.1103/PhysRevLett.104.081301
  [arXiv:0908.2422 [hep-th]].



\bibitem{Pietrykowski:2012nc}
  A.~R.~Pietrykowski,
  Phys.\ Rev.\ D {\bf 87} (2013) no.2,  024026
  doi:10.1103/PhysRevD.87.024026
  [arXiv:1210.0507 [hep-th]].


\bibitem{Anber:2010uj}
  M.~M.~Anber, J.~F.~Donoghue and M.~El-Houssieny,
  Phys.\ Rev.\ D {\bf 83} (2011) 124003
  doi:10.1103/PhysRevD.83.124003
  [arXiv:1011.3229 [hep-th]].

























\end{thebibliography}
\end{document}